\documentclass[prd,onecolumn,nopacs,nofootinbib]{revtex4}

\usepackage{amsmath}
\usepackage{graphicx}
\usepackage{amsfonts}
\usepackage{latexsym}
\usepackage{bbold}
\usepackage{wasysym}
\usepackage{calligra}
\usepackage{float}
\usepackage{ulem}
\usepackage{inputenc}
\usepackage{xspace}
\usepackage{url}
\usepackage{epstopdf}
\usepackage{tikz}
\usepackage{amsthm}

\usepackage{inputenc}

\newcommand{\be}{\begin{equation}}
\newcommand{\ee}{\end{equation}}
\newcommand{\beq} {\begin{equation}}
\newcommand{\eeq} {\end{equation}}
\newcommand{\ba}{\begin{eqnarray}}
\newcommand{\ea}{\end{eqnarray}}

\begin{document}

\title{Non-Riemannian Cosmology: The role of Shear Hypermomentum}

\author{Damianos Iosifidis}
\affiliation{Institute of Theoretical Physics, Department of Physics
	Aristotle University of Thessaloniki, 54124 Thessaloniki, Greece}
\email{diosifid@auth.gr}

\date{\today}
\begin{abstract}
	We consider the usual Einstein-Hilbert action in a Metric-Affine setup and in the presence of a Perfect  Hyperfluid. In order to decode the role of shear hypermomentum, we impose  vanishing spin and dilation parts on the sources and allow only for non-vanishing shear. We then consider an FLRW background and derive the generalized Friedmann equations in the presence of shear hypermomentum. By providing one equation of state among the shear variables we study the cases for which shear has an accelerating/decelerating effect on Universe's expansion. In particular,  we see that shear offers a possibility to prevent the initial singularity formation. We also provide some exact solutions in the shear dominated era and discuss the physical significance of the shear current.
	
\end{abstract}

\maketitle

\allowdisplaybreaks


\tableofcontents

\section{Introduction}

It is widely believed that	non-Riemannian effects, that is torsion and non-metricity, had played a very important and essential role in the very early Universe \cite{puetzfeld2005status}. Studies on these non-Riemannian effects have a long and extended history (see \cite{puetzfeld2005prospects} and references therein). However, almost exclusively in all of these works  some assumption was made regarding the underlying geometry. For instance, almost all of these studies were developed on a Riemann-Cartan, a Weyl or a Weyl-Riemann-Cartan background. That is, a simplified hypothesis on the geometry was made a priori with no discussion on the sources\footnote{As an exception we mention \cite{obukhov1997irreducible}.}. Perhaps this was due to the fact the a Cosmological Model for Hyperfluid with microstructure was lacking so far\footnote{We emphasize here the term Cosmological here. Models for hyperfluids have been around for many years (see for instance \cite{obukhov1993hyperfluid,obukhov1996model,babourova1998perfect}). These however,  could not be regarded as Cosmological since no Cosmological Principle was imposed in their derivation.}. With the recent construction of this kind of model \cite{iosifidis2020cosmological}, the so-called Perfect Cosmological Hyperfluid, it is then natural to concentrate on the irreducible parts on the hyperfluid which are the sources of the non-Riemannian degrees of freedom.

The hypermomentum current can be split into its three irreducible pieces of spin, dilation and shear \cite{hehl1995metric}. This spin part has been studied extensively, and also the dilation is known to be related with scale invariance \cite{Kopczynski:1988jq}. On the other hand, the shear part of the hypermomentum has been illusive so far and its physical relevance is not yet totally clear at the moment. There were some early  attempts to connect it to the hadronic properties of matter \cite{hehl1978hypermomentum}. This possibility seems very interesting indeed but a rigorous construction seems to be missing. In a recent paper \cite{garcia2020projective} it was claimed that the shear current is unphysical and it  should be vanishing. We see therefore that there are contradicting statements  in the literature, regarding the relevance and the physical significance of the shear hypermomentum,  that need to be addressed.
In our opinion the problem is still open and it is exactly the purpose of this work to shed some light into the  concept of shear hypermomentum. To this end, we will start with a Metric-Affine Gravity Theory consisting of the usual Einstein-Hilbert action along with the presence of the Perfect Cosmological Hyperfluid. In order to zero in on the concept of shear we shall impose vanishing spin and dilation parts and only allow the former to be non-zero. More specifically, the paper is organized as follows.  We first set up the formalism  by providing the necessary conventions, definitions etc. Then we briefly review the recently developed Perfect Cosmological Hyperfluid model \cite{iosifidis2020cosmological} along with the relevant components of torsion and non-metricity that are allowed by the Cosmological Principle. Continuing we construct our aforementioned model in a Cosmological setting and by means of the connection field equations we are able to solve for torsion and non-metricity in terms of the sources which are the hypermomentum currents. Moreover, in order to focus exclusively on the role of shear, we impose a vanishing spin and dilation parts and express everything in terms of the shear variables. Then we derive the modified Friedmann equations in the presence of shear hypermomentum along with the relevant conservation laws obeyed by the sources. Finally, by providing an equation of state among the two shear variables we find exact solutions in the shear dominated era (very early universe) and discuss its physical interpretation and significance in Cosmology. In particular we find that shear could prevent the initial singularity for a specific range of the shear 'barotropic' index and also allows the possibility of an accelerating expansion. We also discuss possible constraints on the values of the aforementioned index.

\section{The Geometric  setup}
Let us first start by giving our set of conventions/definition we are going to use throughout this work. We will use the notation of \cite{iosifidis2020cosmological} and therefore we will go rather briefly on the development of the set up here and refer the reader to \cite{iosifidis2020cosmological} for more details.
We shall start with a general non-Riemannian geometry consisting of a metric and an independent affine connection over the manifold $(g,\Gamma,\mathcal{M})$. Our definition for the covariant (on a vector, say) derivative will be
\beq
\nabla_{\mu}u^{\lambda}=\partial_{\mu}u^{\lambda}+\Gamma^{\lambda}_{\;\;\;\nu\mu}u^{\nu}	
\eeq
Our definitions for curvature torsion and non-metricity are
\begin{equation}
R^{\mu}_{\;\;\;\nu\alpha\beta}:= 2\partial_{[\alpha}\Gamma^{\mu}_{\;\;\;|\nu|\beta]}+2\Gamma^{\mu}_{\;\;\;\rho[\alpha}\Gamma^{\rho}_{\;\;\;|\nu|\beta]}
\end{equation}
\beq
S_{\mu\nu}^{\;\;\;\lambda}:=\Gamma^{\lambda}_{\;\;\;[\mu\nu]}
\eeq
\beq
Q_{\alpha\mu\nu}:=-\nabla_{\alpha}g_{\mu\nu}
\eeq
respectively. 	The related contractions of the later two give us the torsion and non-metricity vectors 
\beq
S_{\mu}:=S_{\mu\lambda}^{\;\;\;\;\lambda}\;\;, \;\; Q_{\alpha}:=Q_{\alpha\mu\nu}g^{\mu\nu}\;,\;\; \tilde{Q}_{\nu}:=Q_{\alpha\mu\nu}g^{\alpha\mu}
\eeq
As for the curvature tensor we have the following three ways of contraction
\beq
R_{\nu\beta}:=R^{\mu}_{\;\;\;\nu\mu\beta}\;,\; \hat{R}_{\alpha\beta}:=R^{\mu}_{\;\;\;\mu\alpha\beta}\;\;, \;\; \check{R}^{\lambda}_{\;\;\kappa}:=R^{\lambda}_{\;\;\mu\nu\kappa}g^{\mu\nu}
\eeq
The first one is the Ricci tensor, the second  the homothetic curvature and the last one sometimes is referred to as the co-Ricci tensor. A further contraction gives us the Ricci scalar  which is uniquely defined since
\beq
R:=R_{\mu\nu}g^{\mu\nu}=-\check{R}_{\mu\nu}g^{\mu\nu}\;,\;\; \hat{R}_{\mu\nu}g^{\mu\nu}=0
\eeq
The affine connection can be split into its Riemannian (Levi-Civita) part plus non-Riemannian contributions according to  \cite{schouten2013ricci,iosifidis2019metric}
\beq
\Gamma^{\lambda}_{\;\;\;\mu\nu}=\widetilde{\Gamma}^{\lambda}_{\;\;\;\mu\nu}+N^{\lambda}_{\;\;\;\mu\nu} \label{condec}
\eeq
where 
\beq
N_{\alpha\mu\nu}=\frac{1}{2}(Q_{\mu\nu\alpha}+Q_{\nu\alpha\mu}-Q_{\alpha\mu\nu}) -(S_{\alpha\mu\nu}+S_{\alpha\nu\mu}-S_{\mu\nu\alpha}) \label{l}
\eeq
is called the distortion tensor and $\widetilde{\Gamma}^{\lambda}_{\;\;\;\mu\nu}$ is the usual Levi-Civita connection given by
\begin{equation}
\widetilde{\Gamma}^{\lambda}_{\;\;\;\mu\nu}:=\frac{1}{2}g^{\alpha\lambda}(\partial_{\mu}g_{\nu\alpha}+\partial_{\nu}g_{\alpha\mu}-\partial_{\alpha}g_{\mu\nu})
\end{equation}
Note that quantities with a   tilde will  denote Riemannian parts unless otherwise stated. With the help of the above  decomposition we can express any geometrical object into its Riemannian piece plus non-Riemannian contributions coming from torsion and non-metricity. For instance, substituting the decomposition ($\ref{condec}$) into the definition of the Riemann tensor we deduce
\beq
R^{\mu}_{\;\;\nu\alpha\beta}=\widetilde{R}^{\mu}_{\;\;\nu\alpha\beta} +2\widetilde{\nabla}_{[\alpha}N^{\mu}_{\;\;\;|\nu|\beta]}
+2 N^{\mu}_{\;\;\;\lambda [\alpha}N^{\lambda}_{\;\;\;|\nu|\beta]}
\eeq
In addition,
torsion and non-metricity can be derived from the distortion tensor through the relations \cite{hehl1995metric,iosifidis2019metric},
\beq
Q_{\nu\alpha\mu}=2 N_{(\alpha\mu)\nu} \;\;, \;\;
S_{\mu\nu\alpha}=N_{\alpha[\mu\nu]}
\eeq	
which will be of great use in the sequel.

\section{Cosmology with Torsion and non-metricity}
For the purposes of our analysis we will need the most general acceptable forms of torsion and non-metricity in a homogeneous and isotropic space. It is known that in such a space torsion has at most $2$ degrees of freedom in $4$-dimensions and $1$ for $n \neq 4$ \cite{tsamparlis1979cosmological}. As for non-metricity, there exist $3$ non-vanishing components that are allowed by the Cosmological Principle \cite{minkevich1998isotropic}\footnote{In the case of spherical symmetry the non-vanishing components of torsion and non-metricity were found in \cite{hohmann2020metric}}. As shown in \cite{iosifidis2020cosmological} the latter can be written in an explicitly covariant fashion as
\beq
S_{\mu\nu\alpha}^{(n)}=2u_{[\mu}h_{\nu]\alpha}\Phi(t)+\epsilon_{\mu\nu\alpha\rho}u^{\rho}P(t) \label{torcos}
\eeq
\beq
Q_{\alpha\mu\nu}=A(t)u_{\alpha}h_{\mu\nu}+B(t) h_{\alpha(\mu}u_{\nu)}+C(t)u_{\alpha}u_{\mu}u_{\nu}  \label{Qnmcos2}
\eeq
respectively. Obviously in the above, $A,B$ $C$ represent the non-metric variables while $\Phi$ and $\zeta$ span torsion. In addition, as usual we have used a $1+(n-1)$ spacetime split by introducing the projection tensor $h_{\mu\nu}=g_{\mu\nu}+u_{\mu}u_{\nu}$ with a normalized $n$-velocity field $u_{\mu}u^{\mu}=-1$. The same work we did for torsion and non-metricity can be applied for the distortion tensor and one finds \cite{iosifidis2020cosmological}
\beq
N_{\alpha\mu\nu}^{(n)}=X(t)u_{\alpha}h_{\mu\nu}+Y(t)u_{\mu}h_{\alpha\nu}+Z(t)u_{\nu}h_{\alpha\mu}+V(t)u_{\alpha}u_{\mu}u_{\nu} +\epsilon_{\alpha\mu\nu\lambda}u^{\lambda}W(t)\delta_{n,4}
\eeq
Then, using the defining relations of torsion and non-metricity in terms of the latter we find the relations among their variables
\beq
2(X+Y)=B \;, \;\; 2Z=A\;, \;\; 2V=C \;, \;\; 2\Phi =Y-Z\;, \;\; P = W	
\eeq
or inverting them
\beq
W=P \;, \;\; V=C/2 \;, \;\; Z=A/2	
\eeq
\beq
Y=2\Phi +\frac{A}{2}	\;\;, \;\;\; X=\frac{B}{2}- 2 \Phi -\frac{A}{2}
\eeq
We can therefore parametrize the non-Riemannian effects in Cosmology by either using the set $A,B,C,\Phi, \zeta$ or $X,Y,Z,V,W$. In most practical cases the latter would prove to be more efficient in calculations while the former has more transparent geometrical meaning. In the sequel we shall use a hybrid combination of them. Let us also note that both of the aforementioned sets will be related to the set of sources (hypermomentum) by means of the connection field equations of the Theory.

\section{The Perfect Cosmological Hyperfluid}
Let us know briefly recall the Perfect Cosmological Hyperfluid Model as developed in \cite{iosifidis2020cosmological}. This is the generalization of the Perfect Fluid notion of GR where now the intrinsic properties of matter (i.e spin, dilation and shear) are also taken into account in a Cosmological setup. Accordingly, the forms of the energy-momentum and hypermomentum tensors read
respectively
\beq
T_{\mu\nu}=\rho u_{\mu}u_{\nu}+p h_{\mu\nu}
\eeq 
\beq
\Delta_{\alpha\mu\nu}^{(n)}=\phi h_{\mu\alpha}u_{\nu}+\chi h_{\nu\alpha}u_{\mu}+\psi u_{\alpha}h_{\mu\nu}+\omega u_{\alpha}u_{\mu} u_{\nu}+\delta_{n,4}\epsilon_{\alpha\mu\nu\kappa}u^{\kappa}\zeta \label{hyper}
\eeq
where $\rho$ and $p$ are, as usual, the density and pressure of the perfect fluid component of the hyperfluid and $\phi,\chi,...$ are the variables encoding the non-Riemannian characteristics (hypermomentum microstructure) of the fluid. In the homogeneous Cosmological setup we are considering all of these variables depend mostly on time.  In the above $\delta_{n,4}$ denotes the Kronecker's delta.
The above energy tensors are subject to the conservation laws for the Perfect Cosmological Hyperfluid\footnote{These follow immediately by the diff invariance of the matter part of the action and the assumption that the canonical and the metrical energy momentum tensors of matter coincide \cite{iosifidis2020cosmological}.}
\beq
\tilde{\nabla}_{\mu}T^{\mu}_{\;\;\nu}=\frac{1}{2} \Delta^{\alpha\beta\gamma}R_{\alpha\beta\gamma\nu} \label{ConLaw1}
\eeq
\beq
\hat{\nabla}_{\nu}\Big( \sqrt{-g}\Delta_{\lambda}^{\;\;\;\mu\nu}\Big)=0 \label{ConLaw2}
\eeq
The latter describe the evolution of the perfect fluid and hypermomentum parts of the Perfect Hyperfluid. Of course, in the case where the fluid has no microscopic characteristics ($\Delta_{\alpha\mu\nu}=0$) the latter boil down to the usual conservation law for the  classical Perfect Fluid.

\subsection{ Hypermomentum Split}
As it is well known (see for instance \cite{hehl1995metric}), the hypermomentum current  splits into its three irreducible pieces of spin, dilation and shear according to
\beq
\Delta_{\alpha\mu\nu}=	\Delta_{[\alpha\mu]\nu}+\frac{1}{n}g_{\alpha\mu}D_{\nu}+	\breve{\Delta}_{\alpha\mu\nu}
\eeq
Obviously, the first term on the right-hand side of the above represents the spin part, $D^{\nu}:=\Delta_{\mu}^{\;\;\mu\nu}$ is the dilation and the last one is the shear (traceless symmetric part). 
As  we have already discussed (see \cite{iosifidis2020cosmological}) the most general form of the hypermomentum tensor compatible with the Cosmological Principle (i.e. respecting both isotropy and homogeneity) is given by $(\ref{hyper})$. In this case, the spin, dilation and shear parts read
\beq
\Delta_{[\alpha\mu]\nu}=(\psi-\chi)u_{[\alpha}h_{\mu]\nu}+\delta_{n,4}\epsilon_{\alpha\mu\nu\kappa}u^{\kappa}\zeta
\eeq
\beq
D_{\nu}:=\Delta_{\alpha\mu\nu}g^{\alpha\mu}=\Big[ (n-1) \phi -\omega\Big] u_{\nu} \label{dil}
\eeq
\beq
\breve{\Delta}_{\alpha\mu\nu}=\Delta_{(\alpha\mu)\nu}-\frac{1}{n}g_{\alpha\mu}D_{\nu} =\frac{(\phi+\omega)}{n}\Big[ h_{\alpha\mu}+(n-1)u_{\alpha}u_{\mu} \Big] u_{\nu} +(\psi +\chi)u_{(\mu}h_{\alpha)\nu}
\eeq
respectively. It is the purpose of this article to concentrate on the effects of shear alone (as this has been illusive so far), so in the sequel we shall impose vanishing spin and dilation currents and only keep $\breve{\Delta}_{\alpha\mu\nu}\neq 0$. As a final note before setting up our Theory, let us mention that, as seen from the above discussion,  in $4$ dimensions the spin and shear parts carry $2$ degrees of freedom each and the dilation current is represented by just $1$ degree of freedom.

\section{The Theory}
As we have already discussed our action will consist of the usual Einstein-Hilbert term along with a matter sector encoding the Perfect Cosmological Hyperfluid. That is, our Theory is given by
\beq
S=\frac{1}{2 \kappa}\int d^{n}x \sqrt{-g} R +S_{hyp} \label{S}
\eeq
where 	$S_{hyp}$ is of course the matter part, characterizing the Perfect Cosmological Hyperfluid. Varying the above with respect to the metric and the connection we get respectively
\beq
R_{(\mu\nu)}-\frac{R}{2}g_{\mu\nu}=\kappa T_{\mu\nu} \label{metrf}
\eeq
\beq
P_{\lambda}^{\;\;\;\mu\nu}=\kappa \Delta_{\lambda}^{\;\;\;\mu\nu} \label{conf}
\eeq
where $P_{\lambda}^{\;\;\;\mu\nu}$ is the so-called Palatini tensor, which is given by
\begin{gather}
P_{\lambda}^{\;\;\mu\nu}:=-\frac{\nabla_{\lambda}(\sqrt{-g}g^{\mu\nu})}{\sqrt{-g}}+\frac{\nabla_{\sigma}(\sqrt{-g}g^{\mu\sigma})\delta^{\nu}_{\lambda}}{\sqrt{-g}} 
+2(S_{\lambda}g^{\mu\nu}-S^{\mu}\delta_{\lambda}^{\nu}+g^{\mu\sigma}S_{\sigma\lambda}^{\;\;\;\;\nu})=  \nonumber \\
=	\left( \frac{Q^{\alpha}}{2}+2 S^{\alpha}\right) g^{\mu\nu}-(Q^{\alpha\mu\nu}+2 S^{\alpha\mu\nu})+\left( \tilde{Q}^{\mu}-\frac{Q^{\mu}}{2}-2 S^{\mu} \right)g^{\nu\alpha}
\end{gather}
Some comments are now in order. Firstly, let us note that the above Palatini tensor is traceless in its first two indices ($P_{\lambda}^{\;\;\lambda\mu}=0$) as a consequence of the projective invariance of the Ricci scalar. Such an invariance imposes, by means of ($\ref{conf}$) a vanishing dilation current $\Delta_{\lambda}^{\;\;\lambda\mu}=0$ and therefore restricts the matter form that is allowed to be present. Whether or not projective invariance has a physical significance or not is an open issue, with some recent studies favouring  its existence \cite{jimenez2019ghosts,aoki2019scalar,percacci2019new}. Therefore, in some cases having projective invariance is not fatal but, on the contrary, it  produces healthy Theories. As a result ($\ref{S}$) serves a good starting point for our Cosmological investigation here\footnote{The next logical step would be to add to the Einstein-Hilbert action the $11$ parity preserving quadratic torsion and non-metricity scalars. Of course, this would be considerably more involved and we also have to introduce $11$ free parameters into the Theory (see for instance \cite{obukhov1997irreducible,pagani2015quantum,iosifidis2019scale,jimenez2019general}.) We shall, therefore, leave this for a separate study in the future.}. In addition, since we also impose vanishing dilation and spin currents, whether or not the initial action is projective invariant plays no crucial role and, in fact, even if we had started with a projective breaking version of (\ref{S}) the result would be the same as we show with a specific example in the appendix. Now, going back to our analysis, we shall consider a flat FLRW background with the usual Robertson-Walker line element 
\beq
ds^{2}=-dt^{2}+a^{2}(t)\delta_{ij}dx^{i}dx^{j}
\eeq
where a(t) is as usual the scale factor.  For the rest of the analysis we will use the above metric.

\subsection{Torsion and Non-metricity in terms of their sources (Shear Hypermomentum)}
We now wish to express the torsion and non-metricity variables in terms of their sources (hypermomentum pieces). We will keep our discussion general for the moment and only restrict our attention to the shear part at the end. 
To start with, note that one  can solve for the distortion in terms of the hypermomentum \cite{iosifidis2019exactly} to obtain
\beq
N^{\lambda}_{\;\;\;\mu\nu}=\kappa\frac{g^{\lambda\alpha}}{2}(\Delta_{\alpha\mu\nu}-\Delta_{\nu\alpha\mu}-\Delta_{\mu\nu\alpha})+\kappa\frac{g^{\alpha\lambda}}{(n-2)}g_{\nu[\mu}(\Delta_{\alpha]}-\tilde{\Delta}_{\alpha]})
\eeq
where  $\Delta^{\mu}:= \Delta_{\lambda}^{\;\;\;\mu\lambda}$, $\tilde{\Delta}^{\mu}:= g_{\alpha\beta}\Delta^{\mu\alpha\beta}$
and we have also used the projective freedom to gauge away one vectorial mode \cite{iosifidis2019exactly}. Having the hypermomentum form $(\ref{hyper})$
of the Perfect Cosmological Hyperfluid, we readily compute
\begin{gather}
N_{\alpha\mu\nu}=\frac{\kappa}{2}(\phi -\chi -\psi)h_{\alpha\mu}u_{\nu}+\frac{\kappa}{2}\left[ \frac{1}{(n-2)}(\psi- \chi)-\phi \right] h_{\alpha\nu} u_{\mu}-\frac{\kappa}{2}\left[ \frac{1}{(n-2)}(\psi- \chi)+\phi \right] h_{\mu\nu} u_{\alpha}\nonumber \\
-\frac{\kappa}{2}\omega u_{\alpha}u_{\mu}u_{\nu}-\frac{\kappa}{2}\epsilon_{\mu\nu\alpha\rho}u^{\rho}\zeta \delta_{n,4}
\end{gather}
Comparing this with 
\beq
N_{\alpha\mu\nu}^{(n)}=X(t)u_{\alpha}h_{\mu\nu}+Y(t)u_{\mu}h_{\alpha\nu}+Z(t)u_{\nu}h_{\alpha\mu}+V(t)u_{\alpha}u_{\mu}u_{\nu} +\epsilon_{\alpha\mu\nu\lambda}u^{\lambda}W(t)\delta_{n,4}
\eeq
we find the relation between the distortion and hypermomentum variables
\beq
Y=\frac{\kappa}{2}\left[ \frac{1}{(n-2)}(\psi- \chi)-\phi \right] \;\;,\;\;\; X=-\frac{\kappa}{2}\left[ \frac{1}{(n-2)}(\psi- \chi)+\phi \right]
\eeq
\beq
Z=	\frac{\kappa}{2}(\phi -\chi -\psi) \;\;, \;\;\; V=-\frac{\kappa}{2}\omega \;\; ,\;\;\; W=-\frac{\kappa}{2}\zeta
\eeq
In addition using
\beq
2(X+Y)=B \;, \;\; 2Z=A\;, \;\; 2V=C \;, \;\; 2\Phi =Y-Z\;, \;\; P = W	
\eeq
we can express the torsion and non-metricity functions in terms of their sources(hypermomentum components)
\beq
A=\kappa (\phi-\chi-\psi) \;\;, \;\;\; B=-2\kappa \phi \;\; ,\;\;\; C=-\kappa \omega \label{s}
\eeq
\beq
\Phi=\frac{\kappa}{4}\left[ \frac{1}{(n-2)}\Big( (n-1)\psi +(n-3)\chi \Big) -2 \phi \right]\;\;, \;\;\; P=-\frac{\kappa}{2}\zeta \label{s2}
\eeq
Furthermore, the projective invariance puts the constraint
\beq
(n-1)\phi -\omega=0	
\eeq
We can double check the validity of this result by finding the above connection between the variables in another way. Indeed, given the definition of the Palatini tensor and the forms of torsion and non-metricity for FLRW, we easily compute
\begin{gather}
P_{\alpha\mu\nu}=\left[\frac{(n-3)}{2}A+2(n-2)\Phi -\frac{C}{2}\right] u_{\alpha}h_{\mu\nu}+\left[\frac{(n-2)}{2}B-\frac{(n-1)}{2}A -2(n-2)\Phi -\frac{C}{2}\right] u_{\mu}h_{\alpha\nu} \nonumber \\
-\frac{B}{2}u_{\nu}h_{\mu\alpha}-\frac{(n-1)}{2}Bu_{\alpha}u_{\mu}u_{\nu}-2\epsilon_{\alpha\mu\nu\rho}u^{\rho}P \delta_{n,4}
\end{gather}
Then, using the form of $\Delta_{\alpha\mu\nu}$
\beq
\Delta_{\alpha\mu\nu}^{(n)}=\phi h_{\mu\alpha}u_{\nu}+\chi h_{\nu\alpha}u_{\mu}+\psi u_{\alpha}h_{\mu\nu}+\omega u_{\alpha}u_{\mu} u_{\nu}+\delta_{n,4}\epsilon_{\alpha\mu\nu\kappa}u^{\kappa}\zeta \label{Dform}
\eeq
and the connection field equations
\beq
P_{\lambda}^{\;\;\mu\nu}=\kappa \Delta_{\lambda}^{\;\;\mu\nu}
\eeq
we find the relations
\beq
\frac{(n-3)}{2}A+2(n-2)\Phi -\frac{C}{2}=\kappa \psi
\eeq
\beq
\frac{(n-2)}{2}B-\frac{(n-1)}{2}A -2(n-2)\Phi -\frac{C}{2}=\kappa \chi
\eeq
\beq
B=-2 \kappa \phi
\eeq
\beq
(n-1)B=- 2 \kappa \omega
\eeq
\beq
-2P=\zeta
\eeq
Notice that the latter two imply the projective invariance constraint on the matter variables $(n-1)\phi-\omega=0$ which we have already discussed.
Additionally,  we can fix the gauge (projective freedom) in such a way so as to obtain a vanishing $\tilde{Q}_{\mu}$, then 	$\tilde{Q}_{\mu}=0$
implies 
\beq
C=\frac{(n-1)}{2}B
\eeq	
And upon substituting the latter in the equations above we obtain equivalence with	$(\ref{s})$-$(\ref{s2})$. From these we see that the sources of torsion are only $\psi$ and $\chi$ (and also $\zeta$ for the pseudo-scalar part) while for non-metricity we have all ($\phi,\chi,\psi,\omega$) except $\zeta$ of course. Note that the above derivations were general and no constraints on the hypermomentum (apart from the off shell vanishing dilation) were imposed. 
As we have mentioned in this note we want to unveil the role of shear hypermomentum. To this end we shall impose a vanishing spin part on the hypermomentum, namely
\beq
\Delta_{[\alpha\mu]\nu}=0
\eeq
which implies 
\beq
\psi =\chi \;\;, \;\;\; \zeta=0
\eeq
The above can be thought of as equations os state for the hypermomentum variables. In addition, the vanishing of the dilation current
\beq
\Delta_{\mu}^{\;\;\;\mu\nu}\equiv 0
\eeq
provides another 'equation of state',
\beq
(n-1)\phi-\omega=0 \label{dil}
\eeq 
Using the above constraints, the (shear) hypermomentum is parameterized by two functions of time $\phi(t)$ and $\psi(t)$ and we have
\beq
\Delta_{\alpha\mu\nu}=	
\breve{\Delta}_{\alpha\mu\nu}=\phi \Big[ h_{\alpha\mu}+(n-1)u_{\alpha}u_{\mu} \Big] u_{\nu} +2 \psi u_{(\mu}h_{\alpha)\nu}
\eeq
In this case, the torsion, non-metricity and distortion functions in terms of the sources, are given by
\beq
Y=X=-\frac{\kappa \phi}{2} \;\; ,\;\; Z=\frac{\kappa}{2}(\phi - 2 \psi) \;\;, \;\; V=-\frac{\kappa \omega}{2}\;\;, W=0 \label{X}
\eeq
\beq
\Phi=\frac{\kappa}{2}(\psi -\phi) \;\;, \;\; \zeta=0 \;\;, \;\;	C=\frac{(n-1)}{2}B=-\kappa (n-1)\phi \;\;, \;\; A=\kappa (\phi -2 \psi)  \label{A}
\eeq
Substituting these into the relations for torsion and non-metricity (see eq's ($\ref{torcos}$) and ($\ref{Qnmcos2}$))	we can express the latter in terms of their sources according to
\beq
S_{\mu\nu\alpha}=\kappa (\psi -\phi)u_{[\mu}h_{\nu]\alpha} \label{ts}
\eeq
\beq
Q_{\alpha\mu\nu}=\kappa (\phi-2\psi)u_{\alpha}h_{\mu\nu}-2 \kappa \phi h_{\alpha(\mu}u_{\nu)}-\kappa (n-1)\phi u_{\alpha}u_{\mu}u_{\nu} \label{qs}
\eeq
and we see, as expected,  that we have essentially one degree of freedom for the torsion and two degrees of freedom for non-metricity which are sourced by the two variables $\phi, \psi$ of shear hypermomentum.

\subsection{Conservation Laws}
Let us now turn our attention to the dynamical equations governing the evolution  of the sources of the Perfect Hyperfluid. Expanding the conservation laws ($\ref{ConLaw1}$) and ($\ref{ConLaw2}$) we have
\beq
\dot{\rho}+(n-1)H(\rho+p)=-\frac{1}{2}u^{\mu}u^{\nu}(\chi R_{\mu\nu}+\psi \check{R}_{\mu\nu})\label{rhop}
\eeq
\begin{gather}
-\delta^{\mu}_{\lambda}    \frac{\partial_{\nu}(\sqrt{-g}\phi u^{\nu})}{\sqrt{-g}}-u^{\mu}u_{\lambda}      \frac{\partial_{\nu}\Big(\sqrt{-g}(\phi+\chi +\psi +\omega) u^{\nu}\Big)}{\sqrt{-g}}
\nonumber \\
+\left[ \Big(2 S_{\lambda}+\frac{Q_{\lambda}}{2}\Big)u^{\mu}-\nabla_{\lambda}u^{\mu} \right]\chi +\left[ \Big(2 S^{\mu}+\frac{Q^{\mu}}{2}-\tilde{Q}^{\mu}\Big)u_{\lambda}-g^{\mu\nu}\nabla_{\nu}u_{\lambda}\right]\psi
\nonumber \\
+ u^{\mu}u_{\lambda}(\dot{\chi}+\dot{\psi}) -(\phi+\chi+\psi+\omega)(\dot{u}^{\mu}u_{\lambda}+u^{\mu}\dot{u}_{\lambda}) 
=0   \label{conl2}
\end{gather}
where we have also contracted $(\ref{ConLaw1})$ with $u^{\nu}$. As we have noted, the former is the modified  perfect fluid continuity equation which receives corrections from the hypermomentum, and the latter provide the evolution laws for the hypermomentum currents (i.e. non-Riemannian degrees of freedom). Now, taking the $ij$ components of ($\ref{conl2}$) and also exploiting 	 ($\ref{X}$) and $(\ref{A})$ we find
\beq
\dot{\phi}+(n-1)H\phi +2H \psi=0
\eeq
We should note that taking the $00$ component of the above conservation law will give us an evolution equation for $\omega$ which however provides no new information because of the relation ($\ref{dil}$). It is now natural to assume that there exists a certain equation of state among the shear variables $\phi$ and $\psi$ such that\footnote{This reasoning here is the same with that of the barotropic fluid, where one assumes the usual equation of state $p=w\rho$. } 
\beq
\psi =w_{1}\chi
\eeq
where $w_{1}$ is a constant  which we may refer to as the 'shear' barotropic index. With this, the above conservation law reads
\beq
\dot{\phi}+\Big((n-1)+2 w_{1}\Big)H \phi =0 \label{phic}
\eeq
This is the continuity equation associated with the non-Riemannian degrees of freedom. As we will show in what follows, this very equation will prove to be very crucial and produces many simplifying eliminations making the analysis and the subsequent physical interpretation quite clear. A first illustration of this fact is the following. Computing the right-hand side of ($\ref{rhop}$) and given that  the spin and dilation parts vanish, we find
\beq
u^{\mu}u^{\nu}(\chi R_{\mu\nu}+\psi \check{R}_{\mu\nu})=-(n-1)\kappa \Big( \dot{\phi}+\Big[(n-1)+2 w_{1}\Big]H\phi \Big)
\eeq
but, employing ($\ref{phic}$) the right hand side of the latter vanishes and $(\ref{rhop})$ reduces to the familiar perfect fluid continuity equation
\beq
\dot{\rho}+(n-1)H(\rho+p)=0
\eeq
As a result in this case, the non-Riemannian degrees of freedom decouple from the continuity equation and the perfect fluid component of the hyperfluid evolves in the usual manner. We should mention that this decoupling is by no means a general aspect and only appeared due to the restriction we imposed on the sources. In a general setting, the perfect fluid characteristics mix up with the non-Riemannian ones and the right-hand side of the continuity equation becomes non-vanishing receiving contributions coming from hypermomentum (see \cite{iosifidis2020cosmological}).

\subsection{ Modified Friedmann Equations in the presence of Shear Hypermomentum}
Starting with the metric field equations (\ref{metrf}) and decomposing the Ricci scalar into its Riemannian and post-Riemannian parts (see appendix) taking the $00$ and $ij$ components, after some long calculations we find 
\begin{gather}
H^{2}=-\frac{2}{(n-2)}H\left[ \frac{(n-1)}{2}X-\frac{(n-3)}{2}Y+Z+V \right] -\frac{1}{(n-2)}(\dot{X}+\dot{Y})-\frac{1}{(n-2)}(X-Y)(Z+V)+XY \nonumber \\
+\frac{2}{(n-2)}W^{2}\delta_{n,4}+\frac{2 \kappa}{(n-1)(n-2)}\rho
\end{gather}
\beq
\frac{\ddot{a}}{a}=-\frac{\kappa}{(n-1)(n-2)}\Big[ (n-3)\rho +(n-1)p \Big]+\dot{Y}+H(Y+Z+V)-Y(V+Z)
\eeq
The latter two represent the most straightforward (minimal) generalizations of the Friedmann equations in the presence of torsion and non-metricity. Again, at this point these are fully general in the sense that we have not switched off the spin part yet. Then, using the result of the previous section, if only the shear part is allowed, the above reduce to
\beq
H^{2}=\frac{\kappa}{3}\rho +\frac{\kappa^{2} \phi^{2}}{4}+ \frac{\kappa}{2}\Big( \dot{\phi}+(3+2 w_{1})H \phi \Big) \label{MF1}
\eeq
\beq
\frac{\ddot{a}}{a}=-\frac{\kappa}{6}(\rho + 3 p)-\frac{\kappa^{2} \phi^{2}}{2}(1+w_{1}) -\frac{\kappa}{2}\Big( \dot{\phi}+(3+2 w_{1})H \phi \Big) \label{MF2}
\eeq
where we have also set $n=4$ in order to study physical Cosmology. Notice now that  the last parenthesis on both of the above Friedmann equations	miraculously vanishes by means of ($\ref{phic}$). Then we are left with
\beq
H^{2}=\frac{\kappa}{3}\rho +\frac{\kappa^{2}\phi^{2}}{4} \label{H}
\eeq
\beq
\frac{\ddot{a}}{a}=-\frac{\kappa}{6}(\rho + 3 p)-\frac{\kappa^{2} \phi^{2}}{2}(1+w_{1}) \label{acc}
\eeq
That is, in the end,  there are no linear terms in $H$ and these equations have a more clear physical interpretation. In fact, from the last two equations we see that $\phi^{2}$ acts as an additional perfect fluid component with some peculiar  associated density and pressure which can just as well have a net effect that accelerates the expansion. The above two Friedmann equations are supplemented with the conservation laws ($\ref{rhop}$) and ($\ref{phic}$) and we have a closed system of equations. As final remark let us note that the second modified Friedmann equation $(\ref{acc})$ (also known as the acceleration or Raychaudhuri equation) is not an independent one  and it is a byproduct of the first modified Friedmann equation and the aforementioned two conservation laws. Indeed, as it can be trivially checked, upon differentiating ($\ref{H}$) and employing   both ($\ref{rhop}$) and ($\ref{phic}$) one arrives at  $(\ref{acc})$, as expected. Let us now proceed with the solutions.

\subsection{Solutions}
Let us collect here the system of equations we need to solve in order to unveil the role of shear hypermomentum in Cosmology. The two modified Friedmann equations and the associated conservation laws for the sources, read
\beq
H^{2}=\frac{\kappa}{3}\rho +\frac{\kappa^{2}\phi^{2}}{4} \label{H}
\eeq
\beq
\frac{\ddot{a}}{a}=-\frac{\kappa}{6}(\rho + 3 p)-\frac{\kappa^{2} \phi^{2}}{2}(1+w_{1}) \label{acc}
\eeq
\beq
\dot{\rho}+3H(1+w)\rho =0
\eeq
\beq
\dot{\phi}+(3+2 w_{1})H \phi =0 \label{conw}
\eeq
Before discussing solutions, some comments are in order. Firstly, note  that the effect of shear hypermomentum on the expansion depends crucially upon the sign of $1+w_{1}$ as seen from ($\ref{acc}$). More specifically, for $w_{1}>-1$ shear hypermomentum slows down expansion, for $w_{1}<-1$ it accelerates the expansion and for $w_{1}=-1$ has no effect on the acceleration equation. We shall therefore adopt the suggestive nomenclature and refer to these cases as repulsive ($w_{1}>-1$), attractive ($w_{1}<-1$)  and static ($w_{1}=-1$) shear, respectively. Secondly, from ($\ref{H}$) it follows that shear always enhances the total density and consequently leads to an effective amplification of the latter\footnote{This is also supported by the fact that only $\phi^{2}$ terms appear in the two Friedmann equations meaning that an identification $\rho \propto \phi^{2}$ would be possible. We discuss this in more detail in what follows.}.

\subsubsection{Strong Shear Regime}
It is widely believed (and for good reason) that non-Riemannian effects and in particular those coming from shear hypermomentum had played a significant role at the early stages of the Universe. In particular, for early times the shear degrees of freedom dominate over the perfect fluid characteristics which means that we can safely neglect the latter and focus only on the effect of shear. Indeed, as is also evident from $(\ref{H})$, for early times provided that $3(w-1)<4 w_{1}$, the shear component will dominate and the modified Friedmann equations will reduce to
\beq
H^{2}=\frac{\kappa^{2}\phi^{2}}{4} \label{F1}
\eeq
\beq
\frac{\ddot{a}}{a}=-\frac{\kappa^{2} \phi^{2}}{2}(1+w_{1}) \label{F2}
\eeq
In addition, we can immediately integrate	($\ref{conw}$) to obtain
\beq
\phi= \phi_{0} \left(\frac{a_{0}}{a}\right)^{(3+2 w_{1})}
\eeq
given that for some fixed time $t=t_{0}$ we have $a=a_{0}$ and $\phi=\phi_{0}$. Then substituting the last equation back into  ($\ref{F1}$), we may trivially integrate to arrive at
\beq
a(t)=a_{0} \left[ \pm \frac{\kappa \phi_{0}}{2}(t-t_{0})+1 \right]^{\frac{1}{3+2w_{1}}}
\eeq
for $2w_{1}+3 \neq 0$. The latter describes the cosmological evolution of a shear dominated Universe. We can see clearly that the exact behaviour of the expansion depends on the value of the 'barotropic' shear parameter $w_{1}$. We shall discuss the possible range of values for this parameter in the sequel, based on some general requirements.  It is also apparent from the above expression that in the shear dominated era the growth of the scale factor would be more rapid from that of radiation, given that $w_{1}<-1/2$. Note also that, interestingly,  for $w_{1}=0$ the effect of shear is identical to that of 'stiff' matter.
On geometrical grounds, let us point out that for the specific choice $w_{1}=1$ the whole torsion vanishes as seen from ($\ref{ts}$) while for $w_{1}=1/2$  another one degree of freedom of non-metricity is switched off and only one  survives as can be readily deduced from $(\ref{qs})$.  To close the $3+2w_{1}\neq 0$ case let us also mention that from the two modified Friedmann equations and the definition of the deceleration parameter, one readily finds the latter to be
\beq
q:=-\frac{\ddot{a}a}{\dot{a}^{2}}= 2(1+w_{1}) \label{q}
\eeq
from which again we see that acceleration ($q<0$) is achieved for $w_{1}+1<0$ while for $w_{1}=-1$ we have a 'static'\footnote{We should note that with the term static ($w_{1}=-1$) here refers to shear which has no effect on the acceleration. This is to be distinguished from constant shear $\phi=\phi_{0}$ which as we saw corresponds to the choice $w_{1}=-3/2$.} shear configuration.

For completeness let as also discuss the case for which $3+2w_{1}=0$. As we have already mentioned, in this case the conservation law ($\ref{conf}$) implies $\phi=\phi_{0}=const.$ and the effect of shear will be indistinguishable from that of a Cosmological constant. More precisely, in this case integrating ($\ref{F1}$) we find 
\beq
a(t)=a_{0}e^{H_{0}(t-t_{0})}
\eeq
where we have set $H_{0}=\pm \kappa\phi_{0}/2$. Note that even though $\phi_{0}$ can be either positive or negative, for $3+2w_{1}=0$ it always accelerates the expansion, as seen from the second Friedmann equation, which now reads
\beq
\frac{\ddot{a}}{a}=\left( \frac{\kappa \phi_{0}}{2}\right)^{2}>0
\eeq
Of course, this is due to  the fact that only $\phi^{2}$ appears in ($\ref{F2}$) and there are no linear or other ambiguous sign terms in this equation. In conclusion, for the value $3+2w_{1}=0$ in the shear dominated era the scale factor grows exponentially and with an accelerated manner (i.e. $\ddot{a}>0$). Lastly, as seen from ($\ref{q}$) in this case $q=-1$ as expected.
\subsubsection{Restrictions on the Shear 'barotropic' Index and Physical Interpretation}

On the assumption that the shear part of hypermomentum is related to the hadronic properties of matter \cite{hehl1978hypermomentum} and given the fact that the latter play a significant role in the very early Universe, we can get constraints on the values of $w_{1}$. Indeed, let us fix w, then the shear component would dominate over the perfect fluid contribution in the early Universe, so long as (see eq. ($\ref{MF1}$) )
\beq
w_{1}\geq \frac{3}{4}(w-1) \label{con}
\eeq
Now given the fact that for standard forms of matter $w$ lies in the $[-1,1]$ domain, we obtain for the shear index the inequality $w_{1}\geq-\frac{3}{2}$. The latter condition is also sufficient to ensure that the effect of shear dies out with the passing of time. This can also be inferred from $\ref{phic}$ which for values $w_{1}<-\frac{3}{2}$ would imply that $\phi$ grows larger as the Universe expands and therefore contradicts the assumption of dominance at early times. Therefore, $(\ref{con})$ seems like a logical condition to be imposed on the shear index. Note that an additional constraint may be obtained by demanding that the shear contributions gives always accelerated expansion. Then,  from ($\ref{MF2}$) we must have $w_{1}\leq-1$. The latter assumption being of course, by no means mandatory. In this case $w_{1}$ would be restricted to lie in the interval $[-1,-1/3]$. However, in general it is safe to demand only that $w_{1}$ is greater than or equal to $-3/2$ in order for  shear dominance to be guaranteed in the early Universe.

Let us now try to provide some physical interpretation for the role of shear hypermomentum. To this end we would like to have a perfect fluid component interpretation, that is to relate the shear with some effective perfect fluid characteristics. Interestingly, in our case and given the straightforward form of the modified Friedmann equations, such an association is indeed possible.  To see this, we may multiply $(\ref{phic})$ by $\phi$ and define the shear 'density'
\beq
\rho_{s}:=\frac{3 \kappa}{4}\phi^{2} \label{def1}
\eeq
to get
\beq
\dot{\rho}_{s}+H(\rho_{s}+p_{s})=0
\eeq
where we have  set $w_{s}=1+4(1+w_{1})$ and have also defined the associated shear 'pressure'
\beq
p_{s}:=w_{s}\rho_{s}
\eeq
Note that definition $(\ref{def1})$ as 'density' is possible due to the fact that always $\rho_{s}>0$. On the other hand, the fluids' 'pressure' can be both positive and negative depending on the sign of $w_{s}$. With the above identifications we see that a possible interpretation of the shear effect would be to regard it as an additional perfect fluid  component on top of the regular ($\rho,p$) whose equation of state does not have the restrictions that apply to standard fluids. Indeed, using the above results we may recast the Friedmann equations as
\beq
H^{2}=\frac{\kappa}{3}(\rho+\rho_{s}) 
\eeq
\beq
\frac{\ddot{a}}{a}=-\frac{\kappa}{6}(1 + 3 w)\rho-\frac{\kappa }{6}(w_{s}-1 )\rho_{s}
\eeq
Interestingly, as seen from the last (acceleration) equation above, the existence of shear  speeds up acceleration whenever $w_{s}<1$. We see, therefore, that shear hypermomentum has the same affect as an additional perfect fluid constituent with some exotic equation of state which has also the possibility to accelerate the expansion rather than decelerate it as in the usual case. Here however that 'exotic' fluid is not something unphysical that has been added by hand just to offer more possibilities for the expansion, but has a direct relevance with generalization of the underlying geometry which is now non-Riemannian. Therefore, non-Riemannian effects have very interesting and significant role and affect the Cosmological evolution in a non-trivial way.

As a closing note let us point out that on dropping the assumption that shear is related to the hadronic properties of matter and on the belief that it may just as well have a physical signature on the late time Cosmology, for the choice $w_{1}=-3/2$ the latter acts as a Cosmological constant. Indeed, as we showed earlier in this case, the shear is static and could possibly have late time effects similar to a Cosmological constant.  Of course this is but a speculation and no definite conclusions can be obtained without further analysis. However, in concluding,  we see that the shear current plays a very interesting and essential role in the Universe's evolution and provides some remarkable  new results.

\section{Conclusions}

We have considered the effect of shear hypermomentum in a homogeneous Cosmological setting. Starting with a Metric-Affine action consisting of the usual Einstein-Hilbert term along with the presence of a Perfect Cosmological Hyperfluid, we presented the most minimal generalization of the Friedmann equations in the presence of torsion and nonmetricity. Accordingly, in order to shed light on the role of the shear part of hypermomentum, we imposed  vanishing spin and dilation contributions and only allowed for hypermomentum with non-vanishing shear.  Being equipped with the proper conservation laws for the hyperfluid, we found some exact solutions in the shear dominated era. In particular, the exact effect of shear depends on the equation of state among the shear variables. For specific choices of the shear 'barotropic' index $w_{1}$ the expansion is more rapid then that of radiation and dust. Interestingly, for $1+w_{1}<0$ shear always accelerates expansion and can therefore prevent the initial singularity formation. In addition there exists a certain equation of state (among the  shear variables) for which shear is constant and one obtains accelerated exponential expansion, much like inflation.

In order to give some physical constraint on the shear 'barotropic' index, we demanded that the shear dominated era proceeds that of radiation and dust. This is a logical assumption since it is believed that the shear currents is related to the hadronic properties of matter \cite{hehl1978hypermomentum}  and hadronic currents may play an important role in the early Universe.  One can obtain additional constraints on the 'barotropic' index by demanding for instance that it always leads to accelerated expansion (see equation). We also showed that this type of shear can be seen as a non-standard component of an additional perfect fluid co-existing with the regular one. Interestingly, this perfect fluid component that shear can be mapped to does not have  to obey the restrictions that the usual perfect fluid does. In particular, its associated total Gravitational energy can be just as well negative, leading to an accelerated expansion. Therefore we conclude that the shear current does have a physical relevance and provides new interesting physics for early time Cosmology. Of course, all of the above hold true for the Theory whose Gravitational part consists only by the usual Einstein-Hilbert term and we have also switched off the sources of spin and dilation. The next obvious extension would be to add the quadratic torsion and non-metricity scalars into the Gravitational action and also to switch on the spin and dilation parts of the hypermomentum. This is currently under investigation.

\section{Acknowledgments}
This research is co-financed by Greece and the European Union (European Social Fund- ESF) through the
Operational Programme Human Resources Development, Education and Lifelong Learning in the context
of the project Reinforcement of Postdoctoral Researchers - $2$
nd Cycle $(MIS-5033021)$, implemented by the
State Scholarships Foundation (IKY).

\appendix

\section{Equivalent Theory with Projective Breaking}
We shall now demonstrate that even if we had started with a projective invariant version of $(\ref{S})$ the end result would have been the same due to the fact that we impose a vanishing dilation part. In other words the projective invariance of the Ricci scalar does not cause a problem in this case because the dilation current is switched off anyway. To be more specific let add to $(\ref{S})$ a projective breaking term similar to that suggested in \cite{hehl1981metric}. Our new action reads
\begin{equation}
S=\frac{1}{2\kappa}\int d^{n}x\Big[\sqrt{-g}R+\alpha \sqrt{-g}g^{\mu\nu}\tilde{Q}_{\mu}\tilde{Q}_{\nu}\Big]
\end{equation}
where $\alpha$ is a parameter. Varying the above with respect to the metric tensor and the connection we obtain the field equations
\begin{gather}
R_{(\mu\nu)}-\frac{R}{2}g_{\mu\nu}= \kappa T_{\mu\nu} \alpha \Big[ \frac{1}{2}\tilde{Q}_{\alpha}\tilde{Q}^{\alpha}g_{\mu\nu}-\tilde{Q}_{\mu}\tilde{Q}_{\nu} \nonumber \\
-2 g^{\rho\alpha}(\partial_{\rho}g_{\mu\alpha}g_{\nu\beta})\tilde{Q}^{\beta}-\Gamma^{\lambda}_{\;\;\;\mu\nu}\tilde{Q}_{\lambda}+ \tilde{Q}_{\nu}g_{\mu\alpha}g^{\rho\sigma}\Gamma^{\alpha}_{\;\;\;\rho\sigma}+2 g_{\nu\beta}\frac{\partial_{\mu}(\sqrt{-g}\tilde{Q}^{\beta})}{\sqrt{-g}} \Big] \label{zxcd}
\end{gather}
\begin{equation}
P_{\lambda}^{\;\;\;\mu\nu}+g^{\mu\nu}2 \tilde{Q}_{\lambda}+\delta_{\lambda}^{\nu}2 \tilde{Q}^{\mu}=\kappa \Delta_{\lambda}^{\;\;\;\mu\nu}
\end{equation}
Now, contracting the latter in $\mu,\lambda$ and using the fact that $P_{\mu}^{\;\;\;\mu\nu}=0$ it follows that
\beq
4 \tilde{Q}^{\nu}=\kappa \Delta_{\mu}^{\;\;\;\mu\nu}
\eeq
Imposing a vanishing dilation current, the right-hand side of the above equation vanishes leaving us with $\tilde{Q}^{\nu}=0$. Then substituting the latter back in the field equations, they reduce to
\beq
R_{(\mu\nu)}-\frac{R}{2}g_{\mu\nu}= \kappa T_{\mu\nu}
\eeq
\beq
P_{\lambda}^{\;\;\;\mu\nu}=	\kappa \Delta_{\lambda}^{\;\;\;\mu\nu}
\eeq
which describe the exact system of equations we analyzed in section $V$.

\section{Generalized Friedmann Equations}

In deriving the modified Friedmann equations one needs the components of the Ricci scalar. In order to find them we use the post Riemannian expansion (contracted). In particular, for the $00$ after some algebra we obtain
\beq
R_{00}=-(n-1)\frac{\ddot{a}}{a}+(n-1)H(Y+Z+V)+(n-1)\dot{Y}-(n-1)Y(V+Z)	
\eeq
Then substituting this in the $00$ component of the metric field equations we find the first Friedmann. A similar computation can be done for the $ij$ components but note that this can also be obtained directly from the results of \cite{iosifidis2020cosmic}. There the acceleration equation for the scale factor was derived in full generality with the result being kinematic.

\bibliographystyle{unsrt}
\bibliography{ref}

		\end{document}